# The surface binding and energy issues in rational design of the separation membrane of Li||S batteries

*Shuyu Cheng,[1#] Lijing Wang,[1#] Chao Wu,[2] Sheng Yang,[3] Yang Liu,[1] Yi Zhao,[1] Dandan Cui,[4] Shaowei Zhang,[5] Shixue Dou,[2] Hongfang Du,[1*] Liangxu Lin[1*]*

[1]Strait Institute of Flexible Electronics (SIFE, Future Technologies), Fujian Key Laboratory of Flexible Electronics, Fujian Normal University and Strait Laboratory of Flexible Electronics (SLoFE), Fuzhou, 350117, P. R. China
[2]Institute of Energy Materials Science, University of Shanghai for Science and Technology 200093, China
[3]School of Energy Science and Engineering, Central South University, Changsha, 410083, P. R. China
[4]School of Physics, Beihang University, Beijing 100191, P. R. China
[5]College of Engineering, Mathematics and Physical Sciences, University of Exeter, Exeter EX4 4QF, United Kingdom
*Correspondence: ifelxlin@fjnu.edu.cn (L. L.), ifehfdu@fjnu.edu.cn (H. D.)
[#]Authors contribute equally to this work

**Abstract:** Lithium-sulfur batteries (LSBs) represent one of the most promising next-generation energy storage technologies, offering exceptionally high energy densities. However, their widespread adoption remains hindered by challenges such as sluggish conversion reactions and the dissolution of lithium polysulfides, which lead to poor cycling stability and reduced performance. While significant efforts have been made to address these limitations, the energy storage capabilities of LSBs in practical devices remain far from achieving their full potential. This report delves into recent advancements in the rational design of separation membranes for LSBs, focusing on addressing fundamental issues related to surface binding and surface energy interactions within materials science. By examining the functionalization and optimization of separation membranes, we aim to highlight strategies that can guide the development of more robust and efficient LSBs, bringing them closer to practical implementation.

## 1. Introduction

The lithium-ion battery (LIB), the most successful energy storage device, has reached its theoretical energy density limit.[1] To address the growing demand for higher energy density, lithium-sulfur batteries (LSBs) have emerged as a promising alternative. With a theoretical energy density of 2600 Wh kg$^{-1}$, LSBs offer key advantages such as high safety and the natural abundance of sulfur.[2] This energy density arises from the two-electron redox reaction between lithium (Li) and sulfur (S), supported by the high capacities of the Li anode (3860 mAh g$^{-1}$) and S cathode (1672 mAh g$^{-1}$).[3,4]

Nevertheless, the commercialization of LSBs is hindered by several challenges, including the low conductivity of S, Li dendrite formation, dissolution of lithium polysulfides (LiPSs), and sluggish conversion reactions (**Figure 1a**).[5,6] The poor electrical conductivity of S and its discharge products limits electron transfer and increases electrochemical resistance. Additionally, LiPS dissolution induces the shuttle effect, causing irreversible active material loss. Dissolved LiPSs may convert to short-chain LiPSs or deposit as a passivation layer on the anode, further reducing Coulombic efficiency and cycle stability.[7,8] The formation of Li dendrites exacerbates these issues by consuming electrolytes, reducing Coulombic efficiency, and posing safety hazards.[9] Moreover, during early charge-discharge cycles, reactions between Li and the electrolyte at the solid-liquid interface form a solid electrolyte interphase (SEI) layer. The repeated formation and decomposition of this layer consume electrolytes and degrade the device's capacity over time.[10]

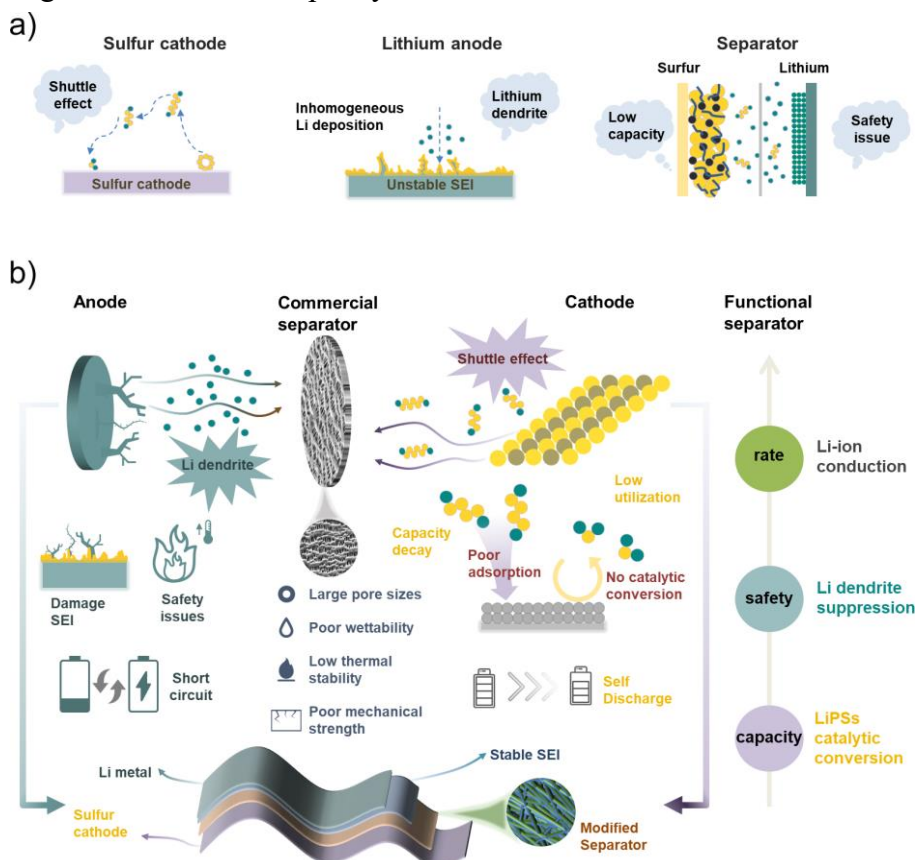

**Figure 1.** a) Challenges facing LSBs. b) Separation membranes in LSBs.

Various strategies have been proposed to address these challenges, including novel S host materials,[11] LiPSs catalysis additives,[12] artificial SEI layers,[13] solid-state electrolytes,[14] and separation membranes.[15] Among these, rationally designed separation membranes are the most effective for mitigating the shuttle effect and suppressing Li dendrite growth. To mitigate the shuttle effect in LSBs, it is crucial to optimize the surface energy of modifying materials and regulate LiPSs binding ability. Effective design must also ensure electrolyte compatibility, enhance lithium ion selectivity, suppress interfacial reactions, and improve mechanical strength to prevent lithium dendrite growth. These interrelated factors significantly impact the performance and stability of the LSBs. Suitable modification of conventional membranes, typically made of single-layer polypropylene (PP), polyethylene (PE), or three-layer PP/PE/PP composites,[16] can also address their weakness of large pore sizes, poor electrolyte wettability, inadequate mechanical strength, and low thermal stability (**Figure 1b**). Additionally, modified separation membranes, closely integrated with the sulfur cathode, can act as "secondary electrodes," facilitating the redox reactions of adsorbed LiPSs, reactivating "dead sulfur," and improving sulfur utilization efficiency. This improvement is achieved through mechanisms such as van der Waals forces, electrostatic attractions, or chemical bonding.[17-20] While previous studies have highlighted the role of separation membranes and their fabrication techniques,[16,21-24] there remains a lack of a unified framework linking separation membrane design to key performance metrics. This review examines innovative materials and structural modifications for separation membranes in LSBs, focusing on surface binding affinity and surface energy issues. By framing these properties as core design principles, we provide a comprehensive overview of current advancements and future directions, underscoring the separation membrane's pivotal role in advancing LSBs technology.

## 2. The surface binding affinity and surface energy

Before delving into specific separation membrane designs, we outline the fundamental roles of separation membranes in LSBs. As highlighted earlier, separation membranes can function as "secondary electrodes". In LSBs, catalytic conversion involves adsorption of reactants, diffusion to active sites, catalysis at active centers, and desorption of products (**Figure 2a**).[25] Optimal catalytic performance requires a moderate binding affinity between the electrode (including the "secondary electrode") and LiPSs.[26-28] However, directly calculating or measuring the binding affinities for all intermediates is challenging. Simplified indicators such as the d/p-band center, coordination number, and bond length are commonly used to estimate binding affinity trends. Notably, for nanoscale systems, finite-size effects may render coordination numbers less accurate, particularly for small nanoparticles (NPs).[29-32]

In principle, the above indicators of d/p-band center, coordination number, and bond length can be linked to the surface energy and "dangling bond" states. As illustrated in **Figure 2b**, unsaturated coordinating atoms fail to fully bond with neighboring atoms, resulting in "dangling bond" states.[33] The density of these dangling bonds serves as a descriptor for surface activity and correlates with surface energy through the following relationship:[34-36]

$$\gamma = \frac{\sum_\alpha I_\alpha^d E_\alpha}{2S}$$

where $\alpha$, $E_\alpha$, $I_\alpha^d$, and $S$ is the bond name, bond energy, equivalent dangling bond number of a bond, and area of the related crystal facets on a particular crystal facet, respectively. Enhancing the dangling bond states can increase the d/p band states and the density of states (DOS) near the Fermi level (**Figure 2b**).[32,33,37-40] This reduces the electron population in antibonding states and strengthens chemical binding. Converting an inactive separation membrane to an active one can improve LiPSs trapping from the electrolyte (facilitating SRR), suppress Li dendrite formation, and enhance the overall SRR. Thus, strategies to improve separation membranes in LSBs can be grounded in surface energy and dangling bond state considerations.

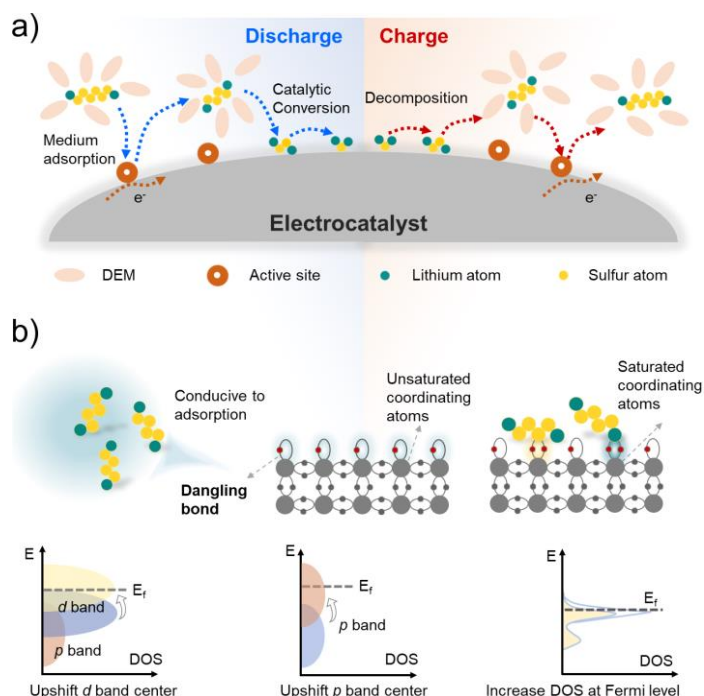

**Figure 2.** a) Schematic of the chemical adsorption and catalytic processes in LSBs. b) Schematics of the dangling bond states in deciding the surface binding affinity, and associated descriptions with the band theory.

## 3. Engineering Strategies

Building on the above concepts, we now explore various strategies to enhance separation membranes in LSBs, including size/morphology control, crystal facet engineering, defect engineering, chemical doping, phase engineering/disordered system, and hybrid structure/surface functionalization.

### 3.1 Size Engineering

The size dependence of materials results from both geometrical and electronic effects. In large crystals, the anisotropic surface electronic structures limit catalytic selectivity.[41] This issue can be alleviated by size reduction. More specially, reducing size, from bulk to NPs and ultimately to single-atom catalysts (SACs), enhances the contribution of corner and edge atoms, increasing surface unsaturation and surface energy (more active sites and improved atomic utilization) compared to bulk materials (**Figure 3a**). This improvement can be simply understood with the change of volume-specific surface area (VSSA) which is inversely proportional to diameter (d) (VSSA = 6/d) of spherical particles.[42]

Thus, the size reduction of active materials can modulate their properties as the separation membrane of LSBs. For instance, the $V_2C$ MXene separation membrane shows increased SRR catalytic activity and prevents Li dendrite formation as its size decreases from 1000 to 150 nm (**Figure 3b**).[43] Want et al. also demonstrated the cage-confined MoC nanoclusters (sub-2 nm) to ensure robust chemisorption for LiPSs and fast SRR.[44] Compared to the above nanomaterials, SACs represent the extreme case of size reduction, offering superior surface-free energy and more unsaturated coordination environments. Quantum effects become significant, leading to discrete electronic energy levels and potential changes in energy band structure as size decreases. In particular, metal SACs can exhibit variable electronic state densities, which influence reaction pathways.[45-47]

Xu et al. demonstrated the use of Ni SACs at MoS$_2$ nanosheets to modify the PP separation membrane (**Figure 3c**), achieving excellent cycling stability (0.01% capacity decay per cycle at 2 C.[48] This configuration also delivered an initial capacity of 6.9 mAh cm$^{-2}$ and 5.9 mAh cm$^{-2}$ after 50 cycles, with a high S loading of 7.5 mg cm$^{-2}$. Other metal SACs such as Fe,[49] Co,[50] Cu,[51] and Mn,[52] have also been explored to modify separation membranes, verifying the potential of SACs in LSBs.

Theoretical understanding on the interaction between various metal SACs to Li-S species can be established through d-p hybridization.[53] As illustrated in **Figure 3d**, the p$_{x/y/z}$ states of S are fully occupied by six electrons, filling low-energy σ and π bonding states. The remaining anti-bonding and non-bonding states are determined by the d orbitals of metal SACs. As the number of 3d electrons increases from Sc to Cu, the d band center lowers away from the Fermi energy level, resulting in more anti-bonding states lower than the Fermi energy level to be filled. This is the same in the band structure of the relative bulk system (**Figure 2b**), where a smaller dangling bond state gives more electron population in the anti-bonding state to reduce the chemical affinity.[33] It is anticipated that transition metals with lower atomic numbers are more efficient in d-p hybridization owing to fewer filled anti-bonding states. Nevertheless, the atomic radius of metals should also be considered to affect the d-p hybridization, as the large atomic radius of Sc weakens the metal-S bond.[53]

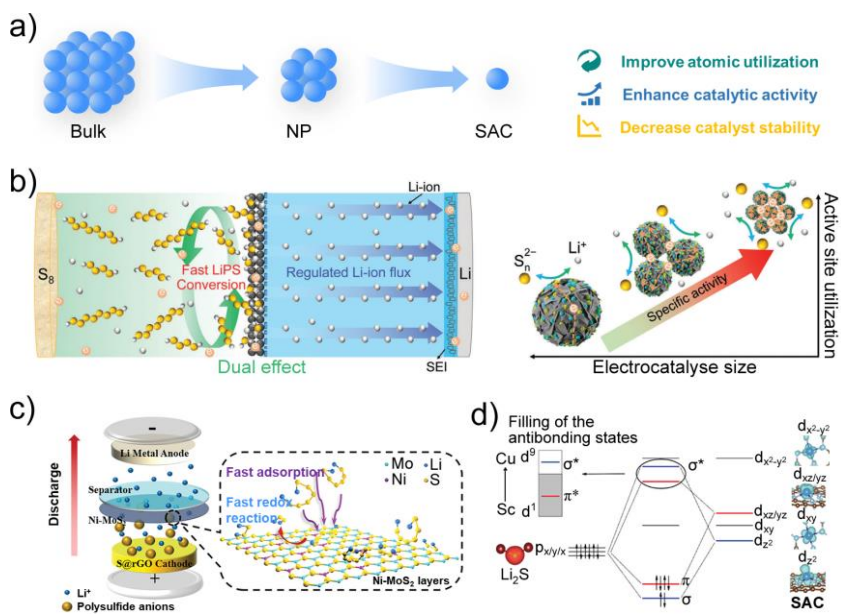

**Figure 3.** a) Schematic of the influences benefited from the size reduction. b) The separation membrane effect and the relationship between the size and electrochemical activity of active material. Reproduced with permission.[43] Copyright 2023, Wiley-VCH. c) Schematic of the working mechanism of Ni-MoS$_2$ nanosheets modified separation membrane in the LSB. Reproduced with permission.[48] Copyright 2023, Wiley-VCH. d) d-p orbital hybridization scenario between SACs and Li$_2$S. Reproduced with permission.[53] Copyright 2021 Wiley-VCH.

Despite these progresses, challenges remain in improving the separation membrane for LSBs:
- The optimal binding affinity of the catalyst for LSBs is still uncertain. While high binding affinity may help trap LiPSs, it is not always beneficial for facilitating fast SRR.[28]

- The actual binding affinities of SACs can differ from theoretical predictions, as these catalysts are typically anchored by external bindings that affect their real interactions with the reactants.
- The high catalytic activity of SACs in the SRR is often compromised by stability issues, including active site detachment and recombination. Excessively strong bonding of LiPSs to the catalytic site can weaken the interaction between the single atom and its support, leading to active site shedding. Furthermore, the dynamic behavior of SACs can result in frequent reorganization, such as aggregation of single atoms into clusters, which diminishes catalytic performance and alters the local electronic environment and reaction selectivity.

In terms of structural stability and activity of SACs, the coordination configuration should be fully considered, aside from the central metal atoms. Various coordinative atoms (e.g., N,[54] O,[55] S,[56] B,[57] and P[58]) have been reported to mount metal SACs. N atom has a high electronegativity (3.04, higher than 2.55 of the C) and a similar size to $Li^+$, therefore, is suitable to interact with the terminal Li atom of LiPSs. Compared to N, the O atom has a higher electronegativity (3.44), giving more charge density transfer to the active site to enhance the stability of SACs, and to provide stronger interaction with $Li^+$.[59] Nevertheless, the above descriptions should be refined, as the N and O doping on carbon host materials are usually performed in different configurations (e.g., in-plane N doping and out-plane doping of O). This is the reason why the N doped configurations are more frequently utilized compared to the O-doped structures. An example here is the M-$N_4$ (M: metal) structure, which has been approved highly efficient in improving structural stability and SRR kinetics.[60,61] These M-$N_4$ sites, such as Mn-$N_4$, have been confirmed capable of weakening the S-S bond in solid $Li_2S_2$ to accelerate the SRR.[62] As for the catalytic property, Fang et al.[63] compared Co-$N_4$ and Co-$N_2$ configurations in performing LSBs. Density functional theory (DFT) simulations suggest that the reduced N coordination of Co-$N_2$ from that of Co-$N_4$ disrupts the symmetric electron distribution, resulting in improved polarity.[63] Consequently, the d-band center of the Co upshifted towards the Fermi level from -0.85 eV of Co-$N_4$ to -0.55 eV of Co-$N_2$, facilitating chemical interactions with LiPSs and charge transfer (also see **Figure 2b**). Similar regulations on the chemical affinity of metal SACs have also been reported elsewhere.[64-66] Nevertheless, this improved chemical affinity also means fewer and weaker bonds of metal SACs to support, reducing the electrochemical stability. Notably, an excessive binding affinity to LiPSs may give "symmetric" SRR pathway with the generation of solid $Li_2S_2$.[28]

The discussions above suggest that reducing the size of active materials in the separation membrane can, in most cases, improve the energy storage performance of LSBs. However, size reduction also increases the instability of the active materials, which must be carefully balanced. In the case of SACs, both the coordination configuration and bonding strength on the support should be considered when evaluating their catalytic properties. It is still challenging in the regulation of the local environment of SACs at the atomic level, particularly concerning the SRR mechanism. It has been suggested that the binding affinity of catalysts should be optimized to prevent the formation of solid $Li_2S_2$,[28] while others argue that an extremely high binding affinity to LiPSs (e.g., in some metal SACs) enhances the dissociation of S-S in $Li_2S_2$, accelerating the SRR.[67,68] The underlying reason may lie in the evolution of the rate-determining step in the SRR, which could involve either preventing the "formation" of $Li_2S_2$ ("asymmetric" SRR),[28] or preventing the "accumulation" of $Li_2S_2$ (S-S dissociation in $Li_2S_2$).[67] Additionally, most reported metal SACs have an atom loading of approximately 0.1-2 wt.%,[69] which should be increased to further enhance catalytic activity. In this context, more theoretical modeling is needed to optimize the coordination configurations of SACs, followed by improved fabrication techniques and enhanced loading ratios.

### 3.2 Crystal Facet Engineering

Crystal facets with a high density of low-coordination atoms possess more dangling bond states, which enhance their chemical affinity to LiPSs and affect the SRR (**Figure 4a**).[70] Fabricating these facet-controlled active materials requires suitable control of surface energy, as facets with lower surface energy tend to form during crystal growth (**Figure 4a**).[70-72] With this in mind, Gong et al. fabricated MnO with (111) facet exposure to modify a carbon-based separation membrane for LSBs.[73] Theoretical calculations indicate that the (111) facet of MnO has a surface energy (γ) of 4.31 J m$^{-2}$, higher than that of the (100) and (110) facets. Similarly, MnO (111) facet, with an adsorption energy ($E_{ads}$) of -11.51 eV, shows the highest chemisorption capacity for $Li_2S_6$, surpassing the (110) facet at -4.41 eV and (100) at -0.94 eV(**Figure 4b**). With this separation membrane, a high capacity (500 mAh g$^{-1}$) after 500 cycles and a low capacity decay rate of 0.05% per cycle was demonstrated.[73] Similarly, Yu et al. incorporated (110)-facet-dominated $VO_2$, with high surface energy, into the separation membrane (**Figure 4c**), achieving an initial capacity of 1472 mAh g$^{-1}$, and excellent cycling performance (1038 mAh g$^{-1}$ remained after 300 cycles at 0.1 C) of the LSB.[74]

This improvement is more evident in high-index facets (HIFs) with higher density of low-coordinated atoms and more dangling bond states over that of low-index facets (LIFs).[75,76] **Figure 4d** illustrates the use of the $SnO_2$ (332) HIF in the LSB's separation membrane.[77] Compared to the $SnO_2$ (111) facet, the $SnO_2$ (332) facet contains more unsaturated threefold- and fivefold-coordinated Sn sites (**Figure 4d**). These Sn sites can adsorb LiPSs and reduce the decomposition energy barrier of solid $Li_2S_2$. DFT calculations suggest that the $SnO_2$ (332) facet has more anti-bonding states above the Fermi level than the $SnO_2$ (111) facet (**Figure 4d**), leading to a correspondingly stronger chemical affinity for LiPSs.

A challenge in facet engineering is the finely controlled growth of crystals, which requires proper utilization of the "surface energy match theory" to minimize the system's total energy.[33] This theory, while not expanded upon here, has been applied effectively in the exfoliation of graphene from bulk graphite.[78,79] These HIFs can be stabilized by ions, surfactants, and through the passivation of surface dangling bond states.[33] **Figure 4a** illustrates the stabilization of HIFs with surface capping agents (e.g., organic molecules, inorganic salts, and polymers)[80-84]. With this strategy, Geng et al. demonstrated the growth of MIL-96-Al MOF with controlled morphology and facet exposure, including hexagonal platelet crystals (HPC), hexagonal bipyramidal crystals (HBC), and hexagonal prismatic bipyramidal crystals (HPBC) (**Figure 4e**).[85] This stabilization of HIFs can also be achieved through the "pinning effect", which involves using an "intermediate" to stabilize two phases (A and B) with significantly different surface energies.[86,87] By incorporating suitable atoms into the HIF, it is possible to balance both stability and activity. Wang et al. demonstrated this technique by alloying Ni and Pt atoms to form (100) facet-exposed surfaces, to improve the chemical affinity for LiPSs and fast SRR kinetics (**Figure 4f**).[88] An extreme example of facet stabilization is high-entropy alloys (HEAs), which offer significantly enhanced mechanical and chemical stability compared to traditional alloys. Thermodynamically, high entropy increases stability through the interactions between constituent elements ($\Delta G = \Delta H - T\Delta S$, where $\Delta G$, $\Delta H$, $T$, and $\Delta S$ are the Gibbs free energy change, enthalpy change, temperature, and entropy change, respectively), especially at high temperatures where the $T\Delta S$ term becomes more significant.[89,90] Kinetically, HEAs also improve structural stability owing to the size mismatch of different elements, which induces lattice distortion and creates large diffusion barriers to prevent phase segregation.[91,92] Xu et al. demonstrated the high electrochemical activity and stability of CoNiFePdV HEA, with a high surface area (308.86 $m^2\ g^{-1}$), even when synthesized at a high temperature (~1300 K) (**Figure 4g**).[93] Similarly, the CoNiCuMnMo-HEA has shown promise as a separation membrane for LSBs, giving a very low capacity decay rate (0.055% per cycle at 3 C).[94] Again, the improved performance of this HEA is ascribed to the enhanced binding affinity to LiPSs.[94]

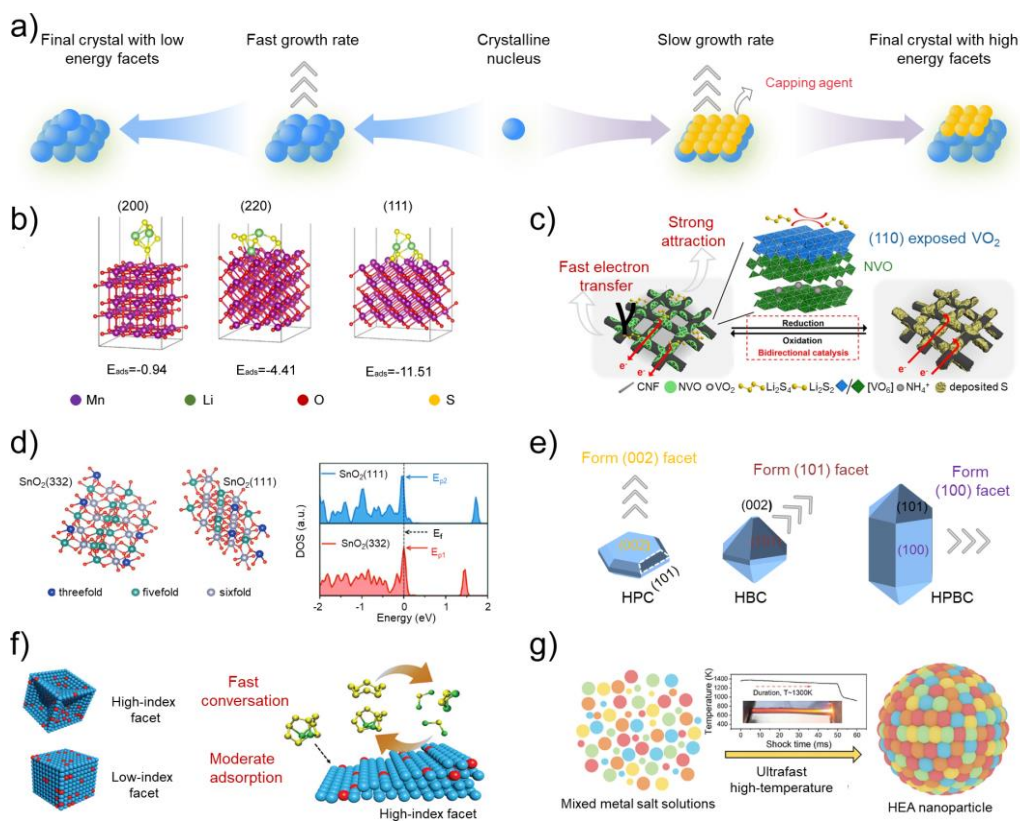

**Figure 4.** a) Schematic of the controlled crystal growth. b) The adsorption model of of $Li_2S_6$ on different crystal facets and corresponding adsorption energies.[73] Copyright 2023, Elsevier. c) Schematic of the bidirectional catalysis of $VO_2$ (110) facet in the LSB. Reproduced with permission.[74] Copyright 2021, American Chemical Society. d) Schematic $SnO_2$ (332) and (111) facets with different unsaturated Sn sites, and the Sn DOS of $SnO_2$ (332) and $SnO_2$ (111) facets. Reproduced with permission.[77] Copyright 2021, Wiley-VCH. e) Diagrams of different crystals. Reproduced with permission.[85] Copyright 2022, Wiley-VCH. f) Schematic of the LiPSs conversion with HIFs and LIFs. Reproduced with permission.[88] Copyright 2022, Wiley-VCH. g) Schematic of HEAs. Reproduced with permission.[93] Copyright 2024, Wiley-VCH.

The above discussions suggest that facet engineering can be well explored in improving the LSBs. With further development in the fabrication and stabilization techniques, we anticipate more promising applications of these facet-engineered materials in LSBs as the separation membrane.

### 3.3 Defect Engineering

Introducing defects into a material can break its structural symmetry, leading to electronic polarization and charge redistribution. These effects are the source of newly formed dangling bond states (or the upshift of the d/p band center, **Figure 2b**) and increased chemical affinity.[33,40,95-99] Defects can take various forms, including vacancies, crystal boundaries, structural dislocations, strained systems, and relaxed bonding structures.[33] Among these, relaxed bonding structures are less commonly discussed, but are significant in small structures. A typical example is the graphene nanoribbon, where the edge structure is metallic, exhibiting enhanced activity and a relaxed bonding structure.[100-103] In defective structures, both the defect site and neighboring atoms can serve as active sites.

Carbon materials have been extensively studied in relation to their electrocatalytic properties in the presence of defects.[104-110] Edge structures (**Figure 5a**), in particular zigzag edges, have much-enhanced chemical affinity to LiPSs compared to that on the basal plane.[111,112] Therefore, to enhance the adsorption of LiPSs, porous carbon materials with a large number of edge sites are better suited for the separation membrane.[111] For example, etched carbon materials with improved edge ratios and microporous structures have been shown to reduce the energy barrier for SRR.[113] However, the microporous structure is not ideal for fast $Li^+$ diffusion. To address this, hierarchical porous carbon may offer a better balance between SRR kinetics and $Li^+$ diffusion.[114] Despite this, the high activity of edge structures, particularly zigzag edges, can lead to structural instability as the edges tend to convert to more stable configurations (e.g., pentagons).[40,97] Similarly, topological defects (e.g., pentagons, heptagons, Stone-Wales defects, and distorted lattices) create asymmetric structures with more dangling bond states, resulting in improved binding affinity for LiPSs.[97,115,116] For instance, carbon membranes with rich pentagonal/heptagonal structures (**Figure 5b**) have shown excellent cycling stability, with a low capacity decay of 0.05% per cycle after 700 cycles at 1 C.[117] One of the main challenges in this field is the stability of carbon materials, which may require additional strategies to enhance electrochemical stability while maintaining high activity for SRR. This issue will be further discussed in **Section 3.4**.

In metal compounds, anion vacancies are commonly introduced to improve chemical affinity. This can be attributed to the resulting unsaturated d-band structure of the metal atom (or the upshift of the d-band center, **Figure 2b**). A typical example is the oxygen vacancy in metal oxides, which has been widely explored.[118-120] Zou et al. reported $In_2O_{3-x}$ NPs with abundant oxygen vacancies, which directly enhanced SRR kinetics by providing both improved chemical affinity and additional channels for $Li^+$ migration.[121] By incorporating the NPs into the separation membrane, a high initial capacity (1540 mAh $g^{-1}$ at 0.05 C) and very low capacity decay (0.058% decay per cycle at 0.2 C) of the LSBs were demonstrated. These effects are also common in 2D transition metal chalcogenides (TMCs) with anion vacancies.[122-128] In 2D $MoS_2$, for example, the unsaturated sulfur vacancies at edges improve binding affinity to LiPSs to promote the SRR (**Figure 5c**).[129]

Besides anion vacancies, cation vacancies can also enhance SRR. To elucidate the critical role of cation defects in SRR, Zhao et al. proposed a mechanism involving the formation of iron (Fe) defects in $Ni_3FeN$.[130] Specifically, Fe atoms are located at the corner sites of the cubic pre-mediator $Ni_3FeN$ (**Figure 5d**). Removal of some Fe atoms through an etching process (**Figure 5d**) results in a surface-distorted $Ni_3Fe_{1-\delta}N$. Because of the asymmetric distribution of electron cloud in Fe vacancies, the d-band center is upshifted to the Fermi energy level to narrow the band gap, which improves binding affinity and significantly accelerates SRR kinetics. However, accurately synthesizing cation defects remains challenging due to their high formation energies, which often require destructive or uncontrolled procedures.[131]

Aside from the above cases, lattice dislocations, strained systems, grain boundaries, and relaxed bonding structures can also be viewed as defective structures.[132-134] As mentioned earlier, improvements in the separation membrane of LSBs can be attributed to the polarized electronic structure (with more dangling bond states) associated with these defects. Han et al. applied the lattice mismatch descriptor of $I_{Latt}$ of TMCs to link the performance in LSBs, where the $I_{Latt}$ equals to $r_{Li-S}/r_{catal}$ ($r_{Li-S}$: lattice spacing of Li-S in LiPSs; $r_{catal}$: lattice spacing of average cations-anions of TMCs).[135] Decreasing the $I_{Latt}$ increases the chemical affinity to LiPSs, which is reached by improving the lattice spacing of TMCs (**Figure 5e**). Again, like that demonstrated in **Figure 2b**, the improved binding affinity to LiPSs is ascribed to the upshift of the d-band center, with more dangling bond states (**Figure 5e**). The scaling relationship confirms that $I_{Latt}$ plays a vital role in activation energies (for the step from $Li_2S_4$ to $Li_2S_2$ of the SRR) and $Li^+$ diffusion (**Figure 5f**), which is suitable for different TMCs. A lower barrier for $Li^+$ diffusion is believed to increase the diffusion rate on the TMC surface, promoting the reaction between lithium and sulfur.[135] Similarly, surface strain can either upshift (tensile strain) or downshift (compressive strain) the d-band center of metal sites (or the p-band in carbon materials), enhancing or weakening the chemical affinity to LiPSs, respectively.[136,137] The relaxed bonding system, similar to tensile lattice strain, increases atomic spacing to expose more dangling bond states for chemical binding with LiPSs. Wang et al. verified these effects by constructing tensile-strained $Ti_3C_2$ MXene/carbon nanotube (CNT) composites.[138] Exerting internal stress on the surface induces lattice distortion and enlarges the Ti-Ti bond, narrowing the energy band and upshifting the d-band center (**Figure 5g**), which improves the binding ability (see **Figure 2b**). Besides, the enlarged energy gap between 3d and 2p orbitals in o-TS-$Ti_3C_2$ facilitates non-metal atoms bonding with other atoms, achieving a higher electronic concentration at the Fermi level, further weakening the S-S bond in LiPSs. Other defects such as grain boundary (GB) can also improve binding affinity based on similar principles, which will not be further extended.[139-141]

It should be noted that the above trend of upshifting the d-band center is only applicable to late transition metals with more than half-filled d-bands. Early transition metals, with less than half-filled d-bands, exhibit lower adsorption energies upon lattice expansion, contrary to the behavior of late transition metals.[142] Specifically, tensile strain leads to reduced bandwidth. In late transition metals, band narrowing increases the population of the d-band, which upshifts the d-band center to maintain d-band filling. In early transition metals, the d-band downshifts to preserve the d-band filling.[142,143]

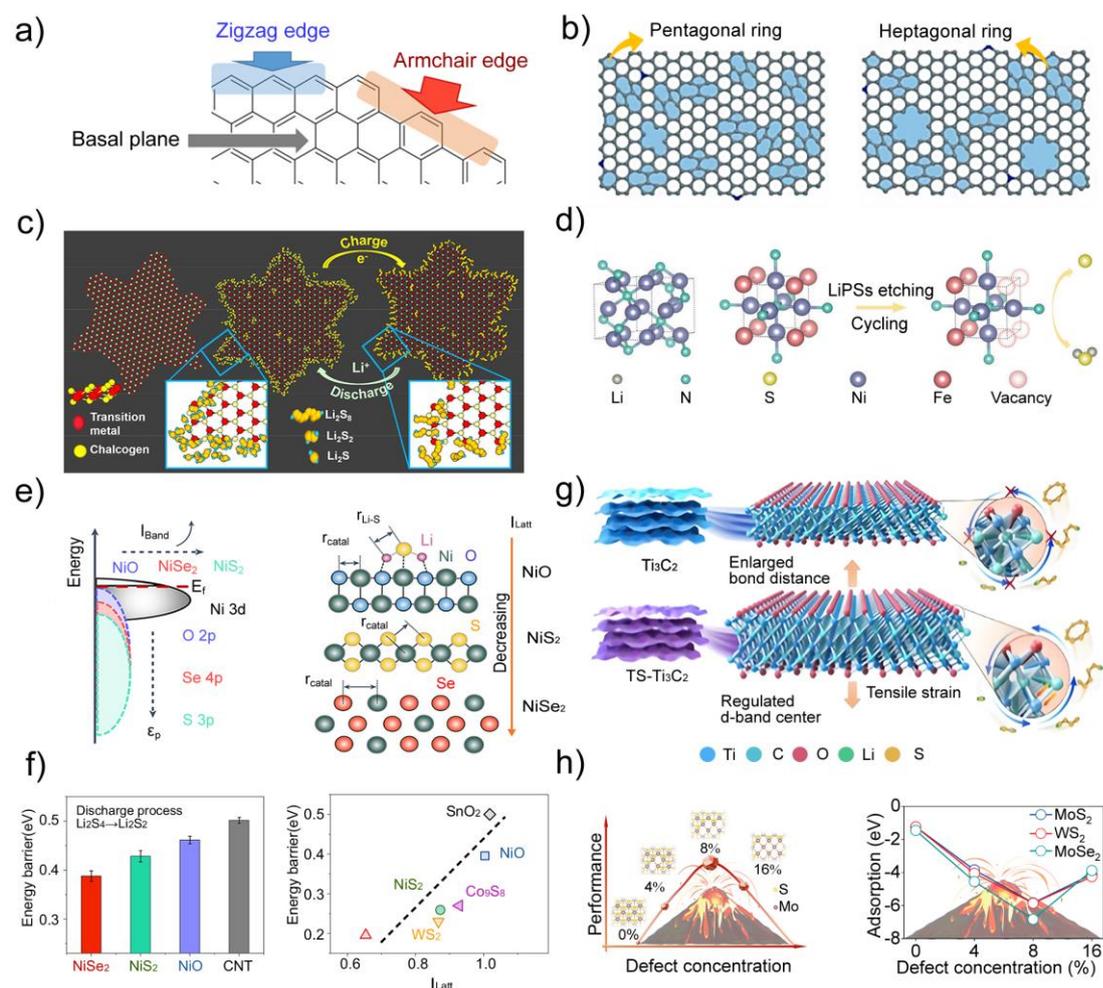

**Figure 5.** a) The basal plane, zigzag and armchair edges of graphene. Reproduced with permission.[112] Copyright 2018, American Chemical Society. b) Graphene with pentagonal/heptagonal rings. Reproduced with permission.[117] Copyright 2023, Springer Nature. c) Schematic of 2D TMCs for LSB, showing confined deposition of LiPSs and their conversions. Reproduced with permission.[129] Copyright 2016, American Chemical Society. d) Structure models of cubic Ni₃FeN and illustration of the LiPSs etching process toward an active Ni₃Fe$_{1-\delta}$N phase. Reproduced with permission.[130] Copyright 2019, Wiley-VCH. e) Schematic of the influence of catalysts with different $I_{Latt}$. f) The relationship between activation energies for Li$_2$S$_4$→Li$_2$S$_2$ (left)/diffusion energy barrier ($E_{barrier}$) (right) and $I_{Latt}$. Figures e, f are reproduced with permission.[135] Copyright 2023, Springer Nature. g) Illustration for the working mechanism of strained Ti$_3$C$_2$. Reproduced with permission.[138] Copyright 2021, Wiley-VCH. h) The volcano relationship between defect level and catalytic performance (left), and plot of binding energy to Li$_2$S$_4$ against defect level of TMCs (right). Reproduced with permission.[145] Copyright 2023, American Chemical Society.

Despite the improvements in LSBs from defect engineering, excessive defects are undesirable, as they may cause catalyst structure collapse and over-binding to LiPSs. For example, in CoP, chemisorption and electrocatalytic activity improve with increasing concentrations of P vacancies,[144] but excessive P vacancies lead to agglomeration or phase transitions of CoP, significantly compromising the stability of the modified separation membrane.[144] To illustrate the relationship between defect concentration and catalytic activity, Guo et al. identified a volcano relationship, showing that only an optimal defect concentration can balance the adsorption and desorption of LiPSs, enhancing the SRR (**Figure 5h**).[145] Another issue associated with defect engineering is the stability of materials. As mentioned earlier, more defects often result in more dangling bond states, which are highly active and may need to be passivated to balance stability and chemical affinity. In this regard, chemical doping can help sustain electronic polarization while maintaining suitable chemical and structural stability, a topic to be further discussed in the next section.

**3.4 Chemical Doping**
The introduction of heteroatoms can induce local electron redistribution and dangling bond states to alter the chemical affinity to LiPSs. Appropriate doping of active materials can also enhance their electrochemical stability. These features make chemical doping a powerful strategy to balance chemical affinity with chemical and structural stability. The following sections will explore non-metal doping (**Section 3.4.1**), metal doping (**Section 3.4.2**), and co-doping (**Section 3.4.3**) of host materials as the separation membrane of LSBs.

3.4.1 Non-metal doping
Both the electronegativity and atomic radius of dopants affect the host materials significantly.[146,147] In carbon host materials, an increase in the electronegativity of dopant relative to carbon leads to a greater accumulation of positive charge on neighboring carbon atoms, thereby enhancing the dipole moment that interacts with LiPSs (e.g., for both $Li^+$ and $S_8^{2-}$). Otherwise, carbon can serve as an active site for adsorbing negatively charged molecules (e.g., $S_8^{2-}$, **Figure 6a**). For negatively charged dopants, the atomic radius should be small enough to facilitate orbital hybridization overlap with Li (excessive atomic radius can induce geometrical distortion and the introduction of O atoms (**Figure 6a**).[147,148] These effects explain why heteroatoms such as N and O with high electronegativity can serve as electron-rich donors, exhibiting a higher affinity for LiPSs (**Figure 6b**).[147-151] For example, introducing an N-doped carbon skeleton into the separation membrane of LSBs results in a very low capacity decay rate of 0.047% per cycle at 1.0 C.[152]

Although F is the most electronegative element, the charge at local binding site is minimal owing to its distinct molecular orbitals. The filled p orbitals of F form a p-π conjugation with carbon, where electron feedback from F to carbon reduces the local charge on F. The combined electron-withdrawing σ effect and electron-releasing p-π effect further reduce the negative charge on F.[147] In contrast, N and O atoms not only participate in the delocalized π system of carbon but also withdraw electrons through σ interactions, which enhances their negative charge. Consequently, F-doped carbon exhibits lower chemical affinity for LiPSs compared to O- and N-doped carbon (**Figure 6b**).[147]

In addition to radius and electronegativity, the presence of a lone pair of electrons in the dopant (serves as a Lewis base) may also affect the chemical affinity of host materials.[147,153] This is exemplified by pyrrolic-N, which has better chemical affinity than other N-doped configurations owing to its lone pair of electrons.[28,154] By incorporating pyrrolic-N configurations into the carbon based separation membrane, Yan et al. demonstrated a stable long-cycle performance of the LSB, giving the reduced energy barrier for the reduction of $Li_2S_4$ and $Li_2S_2$.[155] As we introduced previously in **Section 3.1**, the SRR with N-doped carbon materials is complex and cannot be solely explained by improved chemical affinity for LiPSs. Theoretical modeling shows that the high chemical affinity of pyrrolic-N may lead to the formation of solid $Li_2S_2$, which is undesirable for the fast SRR. This can be avoided by reducing the chemical affinity (note: to structural optimized $Li_2S_4$) slightly on the pyridinic-N and graphitic-C configurations.[155] In contrast, catalysts with much higher chemical affinity (e.g., the binding energy of approximately 6 eV) may enhance the SRR by accelerating the dissociation of the S-S bond in $Li_2S_2$. Discrepancies in the chemical affinity of different N-doped configurations in the literature may be due to improper selection of $Li_2S_x$ configurations.[156,157] Additionally, N-doped carbon materials often suffer from protonation, which may affect electrochemical stability.[158,159] This issue may be not too severe when N-doped carbon materials are solely used as the separation membrane in LSBs.

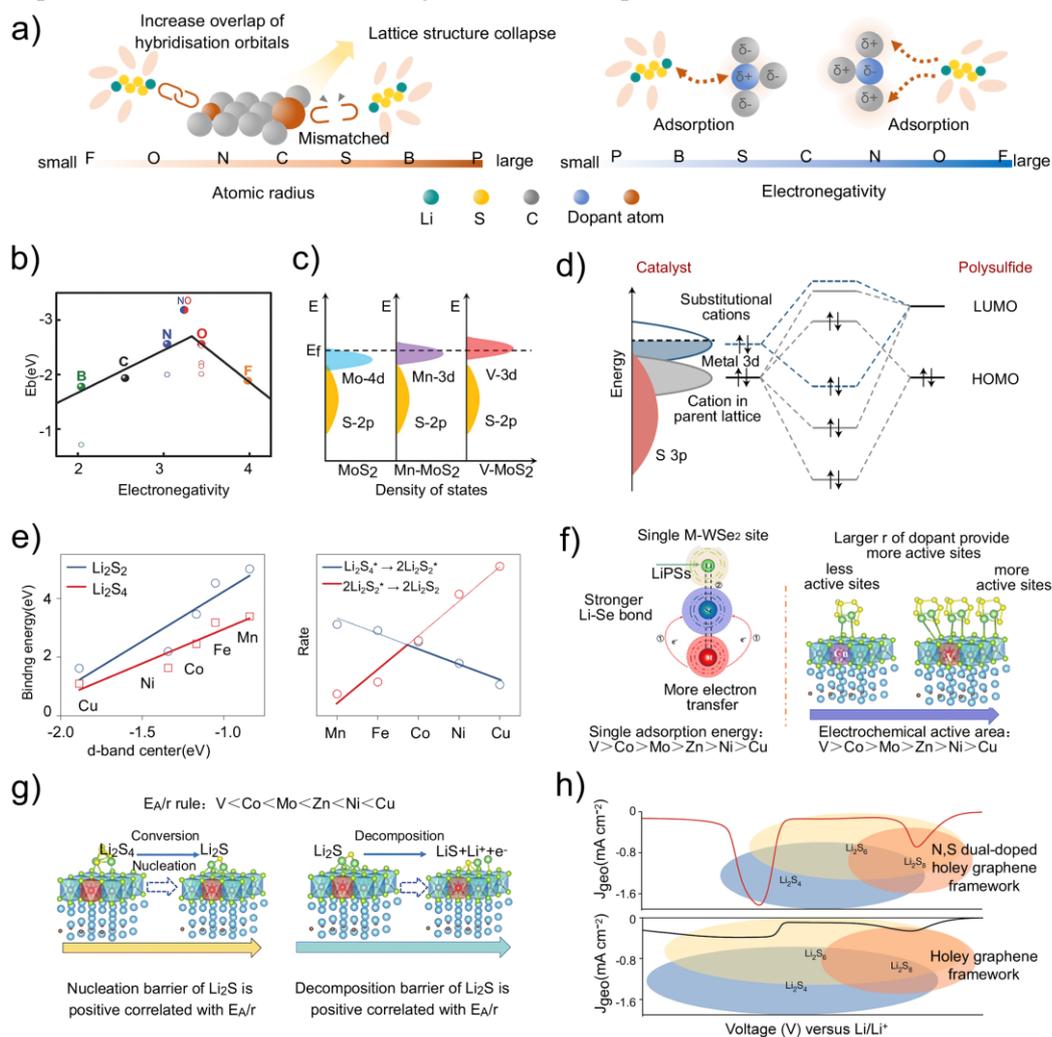

**Figure 6.** a) Schematic of the effect of atomic radius and electronegativity of dopants on carbon materials. b) The binding energy $E_b$ with $Li_2S_4$ versus electronegativity of dopants (carbon for the undoped) of heteroatom-doped graphene nanoribbons. Reproduced with permission.[47] Copyright 2016, Wiley-VCH. c) Conceptual illustration of the d-band shifts upon the chemical doping. Reproduced with permission.[169] Copyright 2023, Wiley-VCH. d) Substitutional cations interact with the lowest unoccupied molecular orbital (LUMO) and highest occupied molecular orbital (HOMO) of polysulfides compared to other cations in parent lattices. e) Relationship of binding energy to d-band center (left) and volcano plots of rates with respect to different dopants (right). Figures d, e are reproduced with permission.[170] Copyright 2022, Springer Nature. f) The relationship between adsorption energy and the electron affinity/ionic radius ($E_{A/r}$). g) The relationship between the nucleation barrier of $Li_2S_4 \rightarrow Li_2S$ and $E_{A/r}$ (left), and the relationship between the decomposition barrier $Li_2S \rightarrow Li_2S_4$ and $E_{A/r}$ (right). Figures f, g are reproduced with permission.[171] Copyright 2023, Royal Society of Chemistry. h) Experimental cyclic voltammetry (CV) curves for N,S-HGF and HGF. Reproduced with permission.[176] Copyright 2024, Springer Nature.

Unlike the N dopant, the B atom cannot afford a lone pair of electrons since it only has three valence shell electrons, giving relatively weak binding affinity to LiPSs of B-doped carbon materials (**Figure 6b**).[147] Nevertheless, the B-doped carbon shares a particular advantage on the electrochemical stability, by considering the proton-like B atom. In this case, a coupling can be formed between the motion of the proton and the distribution of electrons. This coupling improves the electrochemical stability of the catalyst by affecting its electronic structure, proton migration, and interaction energy.[160-164] Consequently, studies have concentrated on co-doping with B and N/O to balance the chemical stability and activity.[165,166]

As for the P and S dopants in carbon materials, their lower electronegativity and larger atomic radius, compared to O and N, result in weaker electronic interactions and longer bond lengths. This leads to reduced electron density regulation and potential structural instability.[157] Therefore, P and S doping alone does not provide satisfactory interactions with LiPSs and is often used in co-doping strategies, which will not be discussed further.[167]

3.4.2 Metal doping
Metal doping significantly influences the Fermi level. In n-type metal doping, where the dopant element has more valence electrons than the host material, the surface dangling bonds readily accept electrons, causing the surrounding atoms to lose electrons. This results in a relative positive charge within the host material, forming an electric field directed toward the surface (upward bending of the surface energy band), and leading to an upshift of the Fermi energy level toward the conduction band (CB).[168] By contrast, in p-type metal doping, where the dopant has fewer valence electrons than the host material, incomplete covalent bonding creates holes. These holes tend to dissociate electrons from the surface, resulting in a positively charged surface and a negatively charged interior. This creates an electric field from the surface to the interior (downward bending of the surface energy band), which downshifts the Fermi energy level toward the valence band (VB).[168] This difference in electronic structure directly impacts the surface chemical affinity for LiPSs.

Liu et al. compared the p-type V-doped $MoS_2$ (V-$MoS_2$) and n-type Mn-doped $MoS_2$ (Mn-$MoS_2$) catalysts as separation membranes in LSBs.[169] The Fermi level upshifted in Mn-$MoS_2$ and downshifted in V-$MoS_2$. The p-type V-$MoS_2$ catalyst exhibited a more pronounced bidirectional catalytic effect, owing to the significant upshift of the d-band center towards the Fermi level and the optimized electronic structure induced by duplex metal coupling (**Figure 6c**). Consequently, the LSBs with the p-type V-$MoS_2$ modified separation membrane delivered excellent long-term cyclic stability (very low decay rate of 0.013% per cycle at 5 C over 5000 cycles). Despite the higher chemical affinity of p-type doping, it does not always outperform n-type doping in terms of long-term cycling stability.[169] Specifically, the presence of holes may lead to the formation of unstable intermediate states, which negatively impact the stability of LSBs under prolonged voltage conditions. In cyclic tests at 1 C, the LSBs with the Mn-$MoS_2$ modified separation membrane showed better stability (capacity decay rate: 0.053% per cycle) compared to the V-$MoS_2$ modified separation membrane (capacity decay rate: 0.069% per cycle).

The above regulations can also be understood from a different perspective. As shown in **Figure 6d**, the original $p_{x/y/z}$ states of sulfur are already occupied by six electrons, filling the low-energy σ and π bonding states. With an increasing number of 3d electrons in the dopant, from Mn to Cu, more anti-bonding states shift below the Fermi level and become filled, leading to weaker interactions with LiPSs.[170] Shen et al. demonstrated the substitutional doping of the ZnS lattice with 3d metal dopants ($Mn^{2+}$, $Fe^{2+}$, $Co^{2+}$, $Ni^{2+}$, or $Cu^{2+}$) to regulate chemical affinity. Notably, the lattice strain induced by these metal dopants also contributed to the shift of the d-band center.[170] **Figure 6e** compares the calculated binding energy of $Li_2S_2$ and $Li_2S_4$ on $M_{0.125}Zn_{0.875}S$ (M: metal dopant). Owing to the maximum elevation of the d-band center, Mn-doped ZnS showed the highest chemical affinity to $Li_2S_2$ and $Li_2S_4$. On the surface of ZnS doped with metal atoms from Cu to Mn, the projected and integrated crystal orbital Hamilton population (pCOHP and iCOHP) of S-S bonds revealed more anti-bonding states away from the Fermi level, weakening the S-S bond. Interestingly, $Mn_{0.125}Zn_{0.875}S$ caused a significant electron redistribution, resulting in a near-zero iCOOP for the S-S bond (note: the smaller the iCOOP value, the weaker the orbital overlap and chemical bonding), indicating that the S-S bond was almost broken. This suggests that $Mn_{0.125}Zn_{0.875}S$ is highly active and can directly dissociate the S-S bonds without the need for electrochemical forces. However, excessive binding between Mn and S could passivate the active site, which explains the volcano trend in the catalytic activity of ZnS doped with metals from Cu to Mn (**Figure 6e**). In this case, the $Co_{0.125}Zn_{0.875}S$ has the highest catalytic activity of SRR.[170]

In metal doping, the ionic radius of the dopant also affects the electronic structure. Want et al. proposed a general electron affinity/ionic radius ($E_A/r$; $E_A$: electron affinity; r: ionic radius of metal dopants) rule as a criterion for selecting metal dopants for $WSe_2$ in LSBs.[171] Among various metal dopants (V, Co, Mo, Zn, Ni, and Cu), a low $E_A/r$ ratio significantly accelerates the SRR. These metal dopants in $WSe_2$ transfer different amounts of electrons into the Se-M (M: metal dopant) bonds, following the order: V (1.194 e) > Co (0.839 e) > Mo (0.640 e) > Zn (0.359 e) > Ni (0.348 e) > Cu (0.335 e). The $E_A$ of these metal atoms has an inverse trend. The low $E_A$ of V leads to a loss of electrons at the metal center, increasing the electron transfer to neighboring Se atoms. This enhanced electron density on Se to improve the binding ability for LiPSs (**Figure 6f**), and to accelerate the breakage or dissociation of LiPSs. This also lowers the energy barrier for $Li_2S$ nucleation and decomposition, promoting fast electrode kinetics (**Figure 6g**). The larger ionic radius of certain dopants can induce more Se vacancies and lattice defects, providing abundant dangling bonds to further enhance the affinity for LiPSs.[171]

3.4.3 Co-doping
Co-doping with non-metallic dopants characterized by high electronegativity (e.g., N or O) and those with large atomic radius (e.g., S or P) improves the overall effectiveness of doping strategies.[172] Dopants with larger atomic radius induce greater lattice distortions, facilitating higher doping levels of highly electronegative elements.[173] The synergistic interaction between these dopants further enhances the affinity for LiPSs and catalytic activity.

For example, the S in the N,S co-doped carbon (NSMPC) promotes the transformation of graphitic-N into pyrrolic- and pyridinic-N, increasing the pyridinic-N content from 38.7% to 40.0% and pyrrolic-N from 38.9% to 42.2%, while decreasing graphitic-N from 12.9% to 9.5%.[174] These pyridinic and pyrrolic-N species, alongside thionic S, co-exist at carbon edges and vacancies, increasing the number of active doping sites.[172,175] As a result, NSMPC showed 148% and 27% improvements in LiPSs adsorption capacity compared to pure and N-doped carbon materials, respectively. When the NSMPC is employed as the separation membrane, the LSBs deliver excellent long-term cycling performance (very low capacity decay: 0.041% per cycle at 0.5 C).[174] Liu et al. further demonstrated the improved SRR kinetics with N, S co-doped holey graphene frameworks (N,S-HGF) (**Figure 6h**).[176] Their study revealed that $Li_2S_6$ is not directly produced by the electrochemical reduction of $Li_2S_8$ but forms via a disproportionation reaction ($Li_2S_8 + Li_2S_4 \rightarrow 2Li_2S_6$) after part of $Li_2S_8$ converts electrochemically into $Li_2S_4$. On N,S-HGF, the conversion of $Li_2S_6$ to $Li_2S_4$ occurred at a higher potential (2.27 V) compared to 2.19 V in undoped HGF. At 1.8 V, the remaining $Li_2S_6$ was only 3% for N,S-HGF, markedly lower than 20% in HGF, indicating accelerated LiPSs conversion. This synergy mitigates the shuttling effect by promoting the rapid depletion of high-order LiPSs at higher potentials.

By combining metal and non-metal doping, interactions with the sulfur and lithium ends of LiPSs can both be improved. Metal dopants introduce impurity energy levels within the CB and VB, acting as donor or acceptor states to regulate electron migration and conductivity.[177-179] Jiang et al. incorporated the P and Mo co-doped $MnO_2$(P,Mo-$MnO_2$) into the separation membrane of the LSB, and demonstrated a high reversible capacity of 927 mAh g$^{-1}$ after 200 cycles at 0.5 C.[179] The P dopant enhances electron donation, increasing the charge density around $MnO_2$'s oxygen atoms.[180,181] Simultaneously, Mo substitutes Mn in the lattice, creating abundant dangling bonds for LiPSs adsorption. This substitution introduces ionized donor states near the Fermi level, lowering the energy barrier for electron excitation to the CB, thus promoting both electronic structure optimization and higher carrier concentration ti improve the chemical affinity for LiPSs[182]

While co-doping of metal atoms can also exploit synergistic effects to modulate electronic and energy band structures, its enhancement mechanisms resemble those of single-metal doping (see **Section 3.4.2**). Thus, a detailed discussion of co-doping is beyond the scope of this section.

### 3.5 Phase Engineering and Disordered System

Phase modulation and the construction of disordered systems can tune electronic properties and surface configurations, enabling precise control over coordination numbers, morphology, and surface arrangements. This strategy leverages increased electronic states near the Fermi level (**Figure 2b**),[125] which has been effectively applied to LSBs.[183-186]

In 2D TMCs, materials can exist in various phases, such as hexagonal 1H/2H, octahedral 1T, and rhombohedral 3R phases.[98] Among these, the 3R phase has received less attention due to its weak interlayer interactions and poor thermodynamic stability. Exfoliation of the 2H phase into 1H monolayers improves surface utilization for electrochemical processes. However, the 1H/2H phases are inherently inert, hindering efficient adsorption of LiPSs. These phases can be transformed into the more active 1T phase. The distinction between the 1H/2H and 1T phases of TMCs is illustrated in **Figure 7a**. For group-VI TMCs, the Fermi level of the 1H/2H phase lies within the energy gap between the $d_{z^2}$ and $d_{x^2-y^2, xy}$ orbitals, giving rise to a semiconducting nature. In contrast, the 1T phase exhibits a Fermi level within the partially occupied d-states ($d_{xy}$, $d_{xz}$, $d_{yz}$), resulting in metallic behavior. This transition makes 1T TMCs more promising for electrochemical processes.[98,125,187,188]

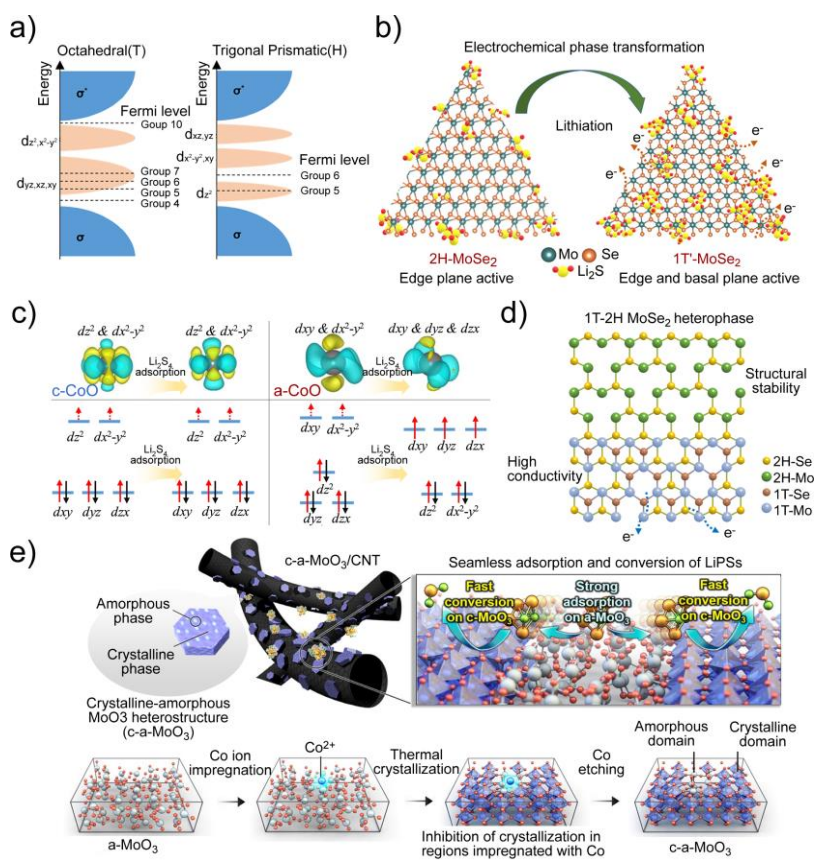

**Figure 7.** a) Schematic of the energy levels of 1T and 1H TMCs with the Fermi-level positions obtained depending on the number of d-electrons. Reproduced with permission.[188] Copyright 2024, Wiley-VCH. b) Schematic of the phase transformation of 2H-MoSe$_2$ into 1T'-MoSe$_2$ upon lithiation and its subsequent influence on the LiPSs adsorption. Reproduced with permission.[192] Copyright 2022, American Chemical Society. c) The calculated deformation change density for Co of c-CoO and a-CoO before (upper left) and after (upper right) Li$_2$S$_4$ adsorption. Reproduced with permission.[204] Copyright 2021, Springer Nature. d) Conversion of 2H phase to 1T-2H heterophase in MoSe$_2$. e) The illustration of crystalline-amorphous MoO$_3$. Reproduced with permission.[213] Copyright 2024, Elsevier.

The 1T phase can further transform into the 1T' and 1T'' phases under structural distortions, improving dangling bond states and surface energy.[189-191] These phases exhibit higher activity (see **Section 3.3**) for LiPSs adsorption and electrocatalytic conversion. **Figure 7b** illustrates the lithiation process that converts 2H-MoSe$_2$ into 1T'-MoSe$_2$[192]. In 2H-MoSe$_2$, only the edge structures are active for LiPSs adsorption. In contrast, both the edge and basal plane of 1T'-MoSe$_2$ are active, doubling the adsorption energy of Li$_2$S$_4$ compared to 2H-MoSe$_2$. The 1T'-MoSe$_2$ also exhibits superior adsorption of Li$_2$S$_n$ (where n ⩾ 4) compared to the dimethyl ether (DME) electrolyte, effectively mitigating the shuttle effect and enhancing SRR efficiency. Additionally, the improved metallicity benefits Li$^+$ diffusion and electron transfer, improving electrode kinetics for SRR. Similar concepts of enhanced metallicity through phase modulation (or improved dangling bond states near the Fermi level) have been demonstrated in MXenes, metal oxides, and metal carbides to improve LSBs.[193-195]

Unlike phase-engineered structures, fully disordered (amorphous) systems lack long-range periodic lattice structures,[196] leading to weak atomic bonds, abundant surface defects, and randomly oriented dangling bonds.[197-200] Amorphization complicates the energy band structure, broadening d-orbital electron distribution and enabling more electrons to reach higher energy levels,[201-203] thus enhancing activity beyond that of phase-engineered materials.[33] Li et al. demonstrated the use of amorphous CoO (a-CoO) in LSBs.[204] As shown in **Figure 7c**, amorphization alters the coordination and symmetry around Co atoms, redistributing the d-orbitals. In crystalline CoO (c-CoO), the high-energy d-orbitals are $d_{z^2}$ and $d_{x^2-y^2}$, while in a-CoO, these shift to $d_{xy}$ and $d_{x^2-y^2}$ orbitals at higher energy levels. This electronic redistribution facilitates electron transfer and enhances chemical affinity for LiPSs. Upon $Li_2S_4$ adsorption, the d-orbitals of a-CoO split into $t_{2g}$ ($d_{xy}$, $d_{xz}$, $d_{yz}$) and $e_x$ ($d_{z^2}$, $d_{x^2-y^2}$) orbitals, promoting greater electron loss and more effective reactions with LiPSs compared to c-CoO.[204] Similarly, amorphous MIL-88B has been reported by Zhang et al. for modifying the separation membrane of LSB, demonstrating exceptional capacity retention of 740 mAh $g^{-1}$ after 500 cycles at 1 C.[205]

While the surface of amorphous structures is highly active, it is inherently unstable. Direct utilization of fully amorphous materials in electrochemical processes is not recommended; they should be passivated or stabilized through structural reconstruction.[33] Metastable 1T TMCs face similar challenges, though they have shown stability in some electrochemical processes. Stabilization of these materials can be achieved by "reducing dangling bonds" and applying the "surface energy matching theory."[33] Reducing dangling bonds through surface passivation improves stability, but may reduce surface activity. Alternatively, ultrathin heterostructures with minimal interlayer energy can enhance the stability of metastable layers.[33] Such structures leverage the synergistic effects of abundant dangling bond states and improve stability during electrochemical processes.[206,207] **Figure 7d** illustrates a 1T-2H $MoSe_2$ heterophase that combines the structural stability of the 2H phase with the high chemical activity of the 1T phase. This heterophase, when incorporated into the separator membrane of LSB, demonstrates excellent rate capability (747.2 mAh $g^{-1}$ at 3 C) and long cycle life (0.079% capacity decay per cycle over 500 cycles).[208] Similarly, heterostructures combining crystalline and amorphous phases can improve the stability of amorphous materials.[209-212] For example, in a crystalline-amorphous $MoO_3$ heterostructure, the amorphous phase promotes robust LiPSs adsorption, while the crystalline phase lowers the energy barrier for $Li^+$ diffusion and $Li_2S$ formation, facilitating SRR (**Figure 7e**).[213]

Although phase engineering and disordered systems can enhance LiPSs adsorption and catalysis, it is crucial to recognize that many highly active materials are thermodynamically unstable.[214] Therefore, balancing stability and high activity in metastable materials is essential.[215] Stabilizing these materials can be achieved through heterostructures with minimal interlayer energy, although the inherent disorder in such systems may reduce controllability and repeatability. This presents challenges for the qualitative investigation of structure-property-performance relationships, potentially limiting further applications.

### 3.6 Hybrid structure and surface functionalization

In hybrid structure and surface functionalized materials, the chemical affinity for LiPSs can also be regulated through effects such as structural coordination and interfacial interaction.[216-237] In hybrid materials combining a high polarity component with a high electrical conductivity component, the polarity component adsorbs LiPSs, while the conductive component accelerates their conversion, enhancing sulfur utilization. In the $MoS_2$-MoN heterostructure (**Figure 8a**), for instance, $MoS_2$ has moderate absorption, mitigating the shuttle effect, while its intrinsic layered structure facilitates fast $Li^+$ diffusion with low energy barriers.[216] MoN, with its high electron conductivity, accelerates the conversion of $Li_2S_x$ (x > 2) to $Li_2S$ and enhances redox kinetics. Yang et al. also proposed MXene@COF hybrid materials as the separator membrane of LSB, demonstrating a low capacity decay rate of 0.048% per cycle after at 1 C.[217] Here, the high surface area and tunable pore structure of COFs facilitate LiPSs adsorption and regulate $Li^+$ flux to suppress Li dendrite growth. In MXene, high electrical conductivity and abundant surface functional groups enhance LiPSs conversion.

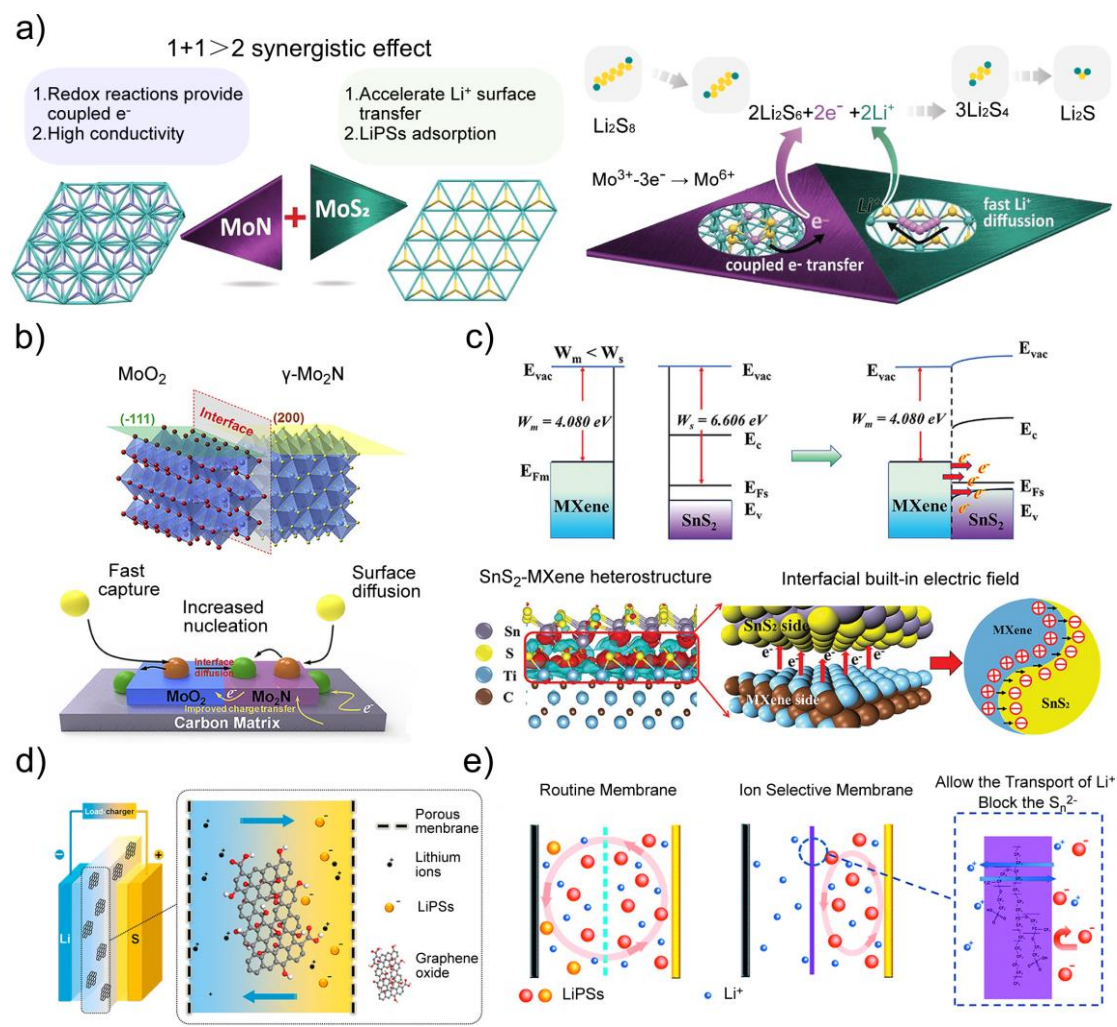

**Figure 8.** a) Schematic of synergistic catalytic conversion of LiPSs by the $MoS_2$-MoN heterostructure. Reproduced with permission.[216] Copyright 2021, Wiley-VCH. b) Schematic of LiPSs conversion and $Li_2S$ nucleation process on the surface of $MoO_2$-$Mo_2N$ heterostructure. Reproduced with permission.[222] Copyright 2020, Elsevier. c) The calculated work function of MXene, $SnS_2$ after contact, and the schematic of electron flow direction simulation in interfacial BIEF. Reproduced with

19 / 40

permission.[223] Copyright 2023, Wiley-VCH. d) Schematic of a GO based separation membrane of LSBs. Reproduced with permission.[231] Copyright 2015, American Chemical Society. e) The LSBs with ion selective membranes. The enlarged local region shows Li$^+$ transport and S$_n^{2-}$ block. Reproduced with permission.[232] Copyright 2014, Royal Society of Chemistry.

In hybrid separation membrane, incorporating polymers or inorganic materials with high elastic modulus can enhance the mechanical strength and stability. This enhancement is demonstrated by the hybrid structure of BN nanosheets and cellulose nanofibers (BNNs@CNFs), where the high elastic modulus and thermal conductivity of BNNs improve heat distribution and suppress dendrite growth.[218] The BNNs@CNFs separator maintains its structure at 150°C for 30 minutes, while the commercial Celgard separator deforms significantly. The high thermal conductivity and elastic modulus of BNNs prevent localized overheating and dendrite penetration, promoting uniform Li dissolution and deposition. Besides, the interfaces within hybrid structures can also enhance electron transfer by introducing interfacial states, energy band alignment, electric field effects, and carrier concentration differences.[219-221] Yang et al. fabricated MoO$_2$-Mo$_2$N hybrid materials as the separation membrane of LSB.[222] MoO$_2$ anchors LiPSs through its strong binding ability, while the high conductivity of γ-Mo$_2$N accelerates LiPSs conversion (**Figure 8b**). The introduction of interfacial states improves charge transport between MoO$_2$ and Mo$_2$N, resulting in excellent cycling stability (capacity decay of 0.028% per cycle at 1 C).

Hybrid structures can also create an electronic polarization in the interlayer, generating a built-in electric field (BIEF) at the interface.[220,221,223-226] The BIEF accelerates Li$^+$ transport and restricts the diffusion of S$_n^{2-}$, suppressing the shuttle effect. It also directs the migration of S$_n^{2-}$ to catalytic sites, facilitating a continuous adsorption-migration-conversion process. For instance, SnS$_2$-MXene Mott-Schottky heterojunctions enhance electron transfer through interfacial BIEF.[223] In this system, Ti atoms in MXene lose electrons, while S atoms in SnS$_2$ gain electrons. Upon contact, electrons flow from Ti to S until thermodynamic equilibrium is achieved (Mott-Schottky effect) (**Figure 8c**). The work function of MXene (4.080 eV) is lower than that of SnS$_2$ (6.606 eV), driving electron transfer from MXene to SnS$_2$, which enhances SnS$_2$'s binding affinity for LiPSs. The heterointerface accelerates Li$^+$/electron transfer, significantly lowering the Li$_2$S nucleation/decomposition barrier and promoting bidirectional sulfur conversion. Additionally, the p-n heterojunction generates a BIEF from n-type to p-type semiconductors, further enhancing the catalytic process in LSBs.[224-226]

Beyond constructing hybrid structures, surface functionalization can modulate surface energy and promote interfacial synergy. Hydroxyl (-OH), sulfonic acid (-SO$_3$H), carboxyl (-COOH), and amino (-NH$_2$) groups are widely used to functionalize separation membranes in LSBs, exploiting effects such as hydrogen bonding, coordination, electrostatic shielding, and spatial hindrance.[227-234] Kong et al. introduced sulfonic acid-functionalized graphdiynes to LSBs separation membranes, reaching a high reversible capacity of 1084.2 mAh g$^{-1}$ and good electrochemical stability of the LSBs (a low attenuation rate of 0.053% per cycle for 1000 long cycles at 2 C).[227] The hydrophilic -SO$_3$H groups increase surface energy, exceeding that of the electrolyte. According to the surface energy matching theory, high-energy surfaces promote better wetting by the lower-energy electrolyte, improving wettability and enhancing ionic conductivity. The strong polarity and spatial hindrance of sulfonic acid groups (S=O, S–

O) effectively inhibit the shuttle effect. Similar improvements have also been demonstrated with the functionalization of the hydroxyl group of active materials.[228-231] **Figure 8d** illustrates the example where carboxyl-functionalized graphene oxide (GO) membranes serve as ion-hopping sites for Li$^+$ while blocking S$_n^{2-}$ through electrostatic repulsion and steric exclusion.[231] More similar regulations can be found on the ion-selective separation membranes with anionic functional groups.[232,233] A cationic selective Nafion membrane with -SO$_3$H groups allows Li$^+$ ion hopping but rejects S$_n^{2-}$ transport through electrostatic interactions (**Figure 8e**). The introduction of the Nafion membrane significantly suppresses the polysulfide shuttle between the cathode and anode.[232,233] Li et al. introduced polydopamine (PDA) and GO to construct an ion-selective separation membrane.[234] The sub-1-nm channels of stacked GO enable selective Li$^+$ migration, while oxygen-containing groups (-COOH and -OH) on PDA prevent S$_n^{2-}$ shuttling via electrostatic interactions. This fully exploits the size-selective and charge-selective effects of membranes. Strong covalent bonding between PDA and GO nanosheets stabilizes channel sizes, ensuring excellent interface stability. The flexible Li-S pouch cell with this separation membrane shows stable performance over 400 cycles and a high practical energy density of 378 W h kg$^{-1}$.

Hybridization and surface functionalization are effective strategies for engineering separation membranes. These strategies affect the electronic structure at the surface or interface to improve surface energy, thus mitigating shuttle effects. However, extremely high surface energy may lead to "over-adsorption" of electrolyte molecules, resulting in uneven electrolyte distribution and compromising battery stability. Additionally, interface stability and electrolyte compatibility must be considered in hybrid materials.

## 4. The Evaluation of Chemical Affinity

The above discussions highlight the crucial role of the chemical affinity of the separation membrane to LiPSs toward more promising LSBs. To ensure efficient capture of LiPSs for catalytic conversions, the chemical affinity of the catalyst to LiPSs should be higher than that of the electrolyte (e.g., > 0.5 eV for DOL and DME electrolytes).[235] As we introduced previously in **Section 3.1**, a high binding may help the dissociation of S-S bond to prevent the accumulation of solid Li$_2$S$_2$ on the catalyst surface.[67,68] Nevertheless, high binding affinity (with the adsorption strength to the geometry optimized Li$_2$S$_4$ as the standard) is not always beneficial to the SRR, as the suitably weaker binding affinity would give asymmetric SRR to avoid the formation of solid Li$_2$S$_4$ intermediate.[28,155] Additionally, excessive affinity of the catalyst surface can result in excessive adsorption of the electrolyte, adversely affecting electrolyte distribution and ion transport. This uneven distribution may increase internal resistance, reduce battery efficiency, and destabilize the solid electrolyte interphase (SEI) film, compromising the safety and stability of LSBs.[236] Therefore, the mechanisms of electrolyte decomposition, such as the reductive decomposition of solvents and lithium salts leading to an excessively thick SEI layer, the side reactions between LiPSs and the electrolyte resulting in surface deposition that hinders ion transport, and the thermodynamic instability at interfaces, remain critical issues that require focused attention. Understanding of the above paradoxical phenomenon should be based on the proper evaluation of chemical affinities to LiPSs, which also forms the fundamentals to guide the proper design of the separation membrane for high-performance LSBs. Therefore, this section will introduce typical protocols to evaluate the chemical affinity of materials to LiPSs, involving experimental methods and theoretical calculations **(Figure 9a)**.

**4.1 Electrochemical Measurement**

Excessive adsorption on the catalyst can lead to passivation, where the surface saturates with reaction intermediates or products such as solid $Li_2S_2$ or $Li_2S$, blocking active sites and hindering further LiPSs reduction. CV measurements with a rotating disk electrode (RDE) can confirm passivation by showing a decrease in current over time, reflecting reduced catalytic activity. The forced convection of RDE ensures a continuous supply of LiPSs, minimizing the impact of concentration changes and providing a more accurate measure of intrinsic activity.

Doping with various cations could enhance the LiPSs affinity of ZnS. Compared with doped samples, the pure ZnS showing the weakest binding ability to LiPSs exhibits the lowest current density in the CV measurement with the RDE technique (**Figure 9b**).[170] In contrast, the $Mn_{0.125}Zn_{0.875}S$, which has the strongest LiPSs affinity, reveals a high initial CV current, indicating high catalytic activity **(Figure 9d)**. Nevertheless, in the low potential range, the current declines gradually over multiple cycles, indicative of the passivation. Notably, reversing the scan to high potential region (>2.2 V) restores the high current in subsequent cycles, indicating that the decay in current density is attributed to passivation rather than catalyst degradation or detachment. As for the $Co_{0.125}Zn_{0.875}S$ with a moderate affinity, it shows consistently high and stable current over cycles **(Figure 9c)**, suggesting minimal passivation. Based on these findings, it is clear that weak chemical affinity is unfavorable for LiPSs adsorption to impede the subsequent conversion, whereas too strong affinity leads to the passivation of catalysts to limit the catalytic activity. Therefore, achieving a moderate chemical affinity is the prerequisite to ensure fast electrode kinetics for rapid adsorption and conversion of LiPSs, which is crucial in the design of efficient separator membranes for LSBs.[170]

The impedance spectroscopy is sensitive to surface adsorption. Consequently, in-situ impedance spectroscopy could reflect how the migration of soluble LiPSs influences charge and mass transfer dynamics in LSBs. The catalyst with higher binding ability typically exhibits reduced impedance changes as running of the battery, indicating a mitigated shuttle of LiPSs. In the case of $Co_{0.12}Ni_{1.88}S_2$/NiO with enhanced chemical affinity, the electrode shows lower and stable electrolyte impedance in situ EIS measurements due to the elimination of the diffusion of LiPSs.[237] In sharp contrast, the NiS/NiO holding inferior affinity exhibits dramatic changes in electrolyte impedance, due to the contamination of LiPSs. As a result, in-situ impedance spectroscopy is a useful yet affordable technique to examine the affinity of LiPSs in the assessment of separator membranes.

**4.2 Spectral Technique**

Ultraviolet-visible spectrometry (UV-vis) can indirectly assess a catalyst's chemical affinity to LiPSs by measuring their concentration in solution, based on the characteristic absorption peaks of LiPSs. By tracking concentration changes of LiPSs before and after the adsorption, the catalyst's affinity can be inferred, as illustrated in **Figure 9e**.[170] However, while this method provides a rough estimate of adsorption efficiency, it cannot precisely determine adsorption energies owing to variations in specific surface areas and active sites among catalysts. A catalyst with a larger surface area or higher density of active sites may show greater adsorption efficiency, even if its intrinsic affinity is weaker. Consequently, UV-vis spectroscopy can rank catalysts based on adsorption performance, but does not offer detailed insights into the strength or nature of the adsorption processes. More advanced techniques, such as in-situ spectroscopy or theoretical calculations, are necessary for a more accurate assessment of the adsorption energies and underlying mechanisms.

In-situ detection techniques, such as Raman and UV-vis spectroscopy, are valuable tools for real-time monitoring of LiPSs. As shown in **Figure 9f-g**, Raman peaks corresponding to LiPSs species (e.g., $Li_2S_4$, $Li_2S_6$) reflect their concentration, with higher concentrations indicating weaker affinity of catalyst.[135] Similarly, in-situ UV-vis spectroscopy tracks LiPSs concentration fluctuations during the SRR by monitoring the absorption peaks that correspond to various LiPSs, providing complementary information on catalyst adsorption behavior **(Figure 9h)**.[238] While both techniques offer insights into the relative adsorption strengths of catalysts, they do not provide precise adsorption energies for further mechanism analysis. This limitation arises from experimental factors such as catalyst surface area, structure, and electrochemical environment. Moreover, these methods can identify trends in adsorption but cannot directly quantify the molecular interactions or binding energies of LiPSs with various catalysts. To better understand the adsorption mechanism and accurately determine catalyst-LiPSs interaction strength, integration with computational methods like DFT is necessary.

In addition, X-ray Photoelectron Spectroscopy (XPS), combined with bond order-length-strength (BOLS) theory, provides a detailed understanding of interface electronic energies and enables the quantitative assessment of adhesive properties at heterogeneous interfaces in two-dimensional materials. At the atomic level, material interfaces are influenced by quantum trapping (QT) or charge polarization (CP), both of which alter the core-level potential of atoms. QT occurs when electron localization at the interface increases surface potential, while CP arises from charge misalignment across the interface. Both mechanisms can affect the strength of interactions between the material and LiPSs.[239]

To quantify these effects, the ratio $\Delta E_v(I)/\Delta E_v(B) = \gamma$ is used, where $\Delta E_v(I)$ and $\Delta E_v(B)$ represent the energy transfers at the sulfide atomic interface and bulk binding energy shifts, respectively.[239] A $\gamma > 1$ indicates QT dominance, leading to weaker interactions, while $\gamma < 1$ suggests CP dominance, enhancing interactions. The interfacial adhesive energy ($\Gamma$), which governs catalyst binding with LiPSs, is calculated using BOLS theory's bond energy and length formulas. The greater the atomic cohesive energy ($E_c$), the better the stability of the material **(Figure 9i)**. For materials such as CS, CuS, $FeS_2$, $MoS_2$, and $SnS_2$, the $\Gamma$ are 2.33, 3.42, 5.97, 5.60, and 4.69 $J/m^2$, respectively. $FeS_2$ exhibits the strongest binding ability, followed by MoS$_2$, SnS$_2$, and CuS. These results indicate that materials with higher adhesive energies, like FeS$_2$, are more effective in anchoring LiPSs and reducing the shuttle effect.[239] These insights, which are difficult to obtain with conventional techniques, make XPS and BOLS theory powerful tools for optimizing electrode materials in LSBs. By elucidating the fundamental mechanisms of catalyst-LiPSs interactions, these methods can inform the design of materials with enhanced catalytic properties and improved long-term stability.

### 4.3 Theoretical Calculation

DFT calculations and molecular dynamics (MD) simulations provide critical insights into the atomic-level interactions between catalysts and LiPSs in LSBs. These techniques enable precise calculation of binding energies, essential for understanding how materials adsorb and stabilize LiPSs. DFT and MD simulations offer theoretical values for chemical affinities, guiding the design of electrode materials with improved performance. For example, DFT calculations by Li et al. directly show that CoN$_4$-carbon surfaces offer stronger adsorption energy toward Li$_2$S and Li$_2$S$_6$ than pure carbon or N-doped carbon.[240]

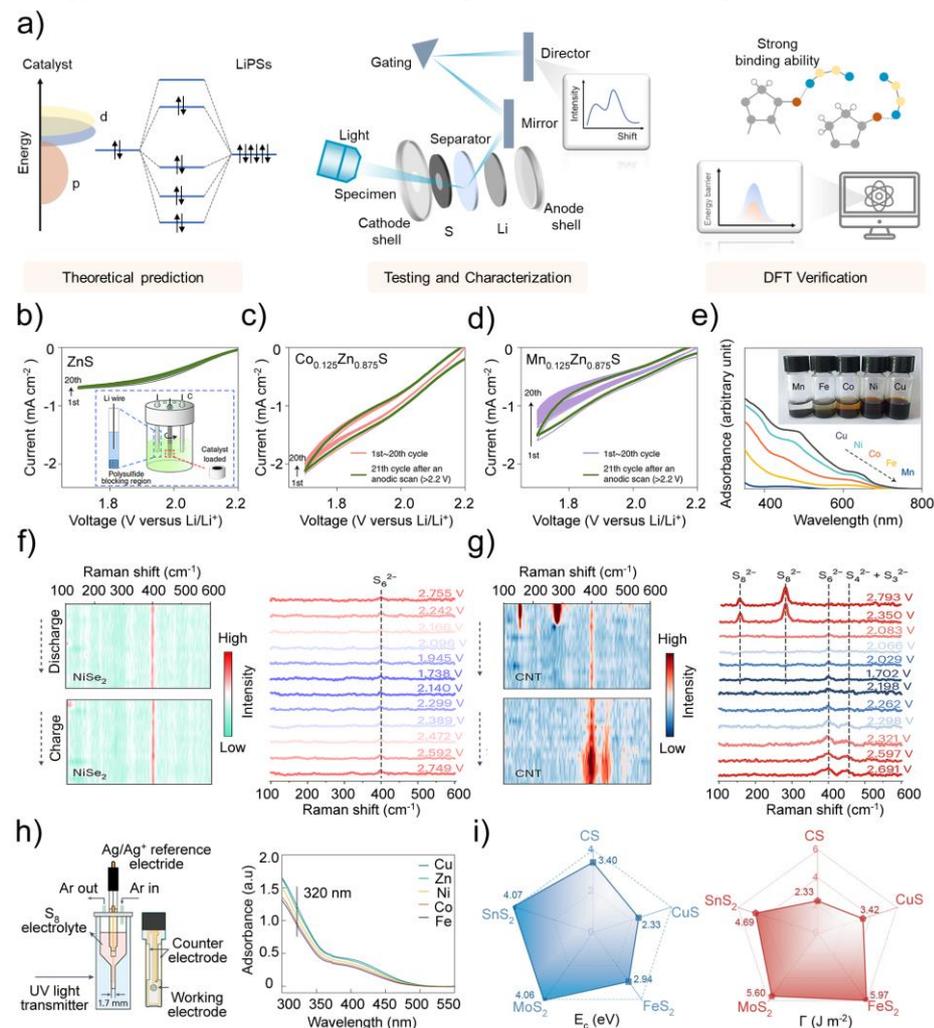

**Figure 9.** a) Typical steps in evaluating affinity include: theoretical predication, experimental testing, and DFT verification. CV curves of ZnS (b), Co$_{0.125}$Zn$_{0.875}$S (c) and Mn$_{0.125}$Zn$_{0.875}$S (d) at 50 mV s$^{-1}$ and 1,000 r.p.m. The inset in d is an illustration of the RDE setup using a Li reference electrode. e) UV-vis absorption spectra and visualization tests of the interaction between LiPSs and M$_{0.125}$Zn$_{0.875}$S catalysts. Figures b-e are reproduced with permission.[170] Copyright 2022, Springer Nature. In situ Raman spectra of the electrode with (f) and without (g) NiSe$_2$ during charging and discharging at 0.3 C. Figures f, g are reproduced with permission.[135] Copyright 2023, Springer Nature. h) Schematic of the in situ UV-vis cell set-up used to quantify LiPSs concentrations during the SRR and the UV-vis curves for different catalysts at -1.1 V versus an organic Ag/Ag$^+$ reference electrode. Reproduced with permission.[238] Copyright 2024, Springer Nature. i) The atomic cohesive energy EC and the interfacial adhesion energies Γ for CS, CuS, FeS$_2$, MoS$_2$, and SnS$_2$ in the interface region. Reproduced with permission.[239] Copyright 2022, Springer.

Nevertheless, DFT calculations primarily focus on equilibrium states rather than the kinetics or dynamic processes involved in reactions, like the shuttle effect. While DFT can provide insights into reaction mechanisms, transition states, and activation energies, it does not capture real-time dynamics such as adsorption/desorption rates or reaction kinetics. Methods like MD or kinetic Monte Carlo (KMC) simulations are better for these tasks.[241,242] Additionally, DFT neglects solvent effects which are crucial for LiPSs behavior in the electrolyte, necessitating simulations that incorporate solvation effects or more complex models.

The integration of electrochemical analysis, spectroscopic evaluation, theoretical modeling, and in-situ monitoring is expected to provide a more comprehensive and accurate understanding of catalyst surface chemistry. As these methods advance, it may become possible to directly obtain rigorous binding energy data through mathematical and physical approaches. This development could greatly inform experimental design, fostering mutual validation and complementary improvements across research efforts.

## 5. Summary and Perspectives
### 5.1 Summary of The Surface Binding Affinity and Engineering Techniques
This report provides an in-depth review of recent advances in the rational design of separation membranes for LSBs, focusing on addressing fundamental issues related to surface binding and surface energy interactions within materials science. Optimizing the surface energy and increasing dangling bond states of separation membranes can enhance the adsorption and catalysis of LiPSs, thereby facilitating the SRR and reducing lithium dendrite formation. The following key insights have been derived:

- **Relationship between chemical affinity and catalytic activity**: The moderate chemical affinity of separation membranes is crucial for effectively trapping LiPSs and fast conversion reactions. Lack of affinity results in weak bonding ability with LiPSs. Excessive affinity may lead to catalyst poisoning, hindering conversion reactions. This highlights the need for balanced adsorption strength to prevent the detrimental accumulation of byproducts.
- **Electrolyte competition for adsorption**: Electrolytes (e.g., DOL, DME) can compete LiPSs adsorption on the separation membranes. The membrane's affinity for LiPSs must surpass that of the electrolyte to enable selective capture of LiPSs. Excessive affinity may lead to over-adsorption of electrolytes, disrupting ion distribution and transport. This imbalance increases internal resistance and destabilizes the SEI film, compromising the safety and stability of LSBs.
- **Instability of high surface energy states**: Separation membranes with high surface energy may exhibit increased activity owing to abundant dangling bond states, however, they are unstable. Strategies for passivating high-energy sites, such as surface passivation, use of additives, and post-processing techniques, are crucial for balancing activity and durability, thereby enhancing long-term performance and preserving catalytic efficiency.

- **Complexity of structure configuration**: The optimal binding affinity for LiPSs is influenced by the configuration of active sites, such as pyrrolic and pyridinic N. Balancing different configurations and binding ability is essential to prevent excessive trapping of LiPSs, while ensuring efficient conversion and preventing the accumulation of solid $Li_2S_2$ on the catalyst surface.
- **Selection of the most stable configuration in DFT calculations**: Given these intricate interactions, computational methods, particularly DFT calculations, are essential for accurately predicting and optimizing binding affinities across various configurations. DFT calculations provide insights into how geometric structures and surface energies influence the adsorption and catalytic behavior of LiPSs. However, selecting the most stable computational model is critical, as minor variations in the calculation method can result in significant discrepancies in predicted binding energies.

These insights highlight the critical balance between surface energy, surface affinity, and material stability in the design of separation membranes for LSBs. Optimizing these parameters, along with selecting the appropriate active site configurations, is essential for enhancing LiPSs adsorption, accelerating the SRR, and ensuring long-term membrane performance. Future research should focus on fine-tuning these factors to achieve a stable, high-performance membrane that effectively mediates LiPSs conversion while maintaining material durability.

To enhance the performance of separation membranes in lithium-sulfur batteries, various modification strategies have been explored, including size engineering, crystal facet engineering, defect engineering, chemical doping, phase engineering and disordered systems, hybrid structures, and surface functionalization. These strategies aim to optimize the properties of separation membranes, such as surface energy, active site configuration, and structural integrity. The key principles behind these approaches are summarized as follows:
- **Single-atom catalysts (SACs):** SACs enhance the performance of separation membranes in LSBs, offering increased surface energy and more unsaturated coordination sites to improve SRR activity. However, optimizing the binding affinity and stability of SACs remains challenging, particularly in preventing excessive binding to LiPSs and the formation of solid $Li_2S_2$. Further theoretical and experimental work is needed to refine SAC fabrication and enhance their catalytic performance in separation membranes.
- **High-index facets (HIFs):** HIFs can provide a higher density of low-coordinated atoms, improving the binding ability to LiPSs. Surface passivation, alloying, and HEA are reported to further balance the stability and catalytic activity of these facets. These facet-engineered materials show significant potential for improving SRR kinetics and long-term stability in LSBs.
- **Defective structure:** Defective structures can introduce electronic polarization and increase the number of dangling bond states to improve the binding affinity for LiPSs. Defects like vacancies, dislocations, and strained systems create unsaturated sites that enhance interactions with LiPSs, accelerating the SRR. However, excessive defects can cause structural instability and over-binding to LiPSs, compromising performance. Thus, controlling defect concentration is vital for balancing catalytic activity and stability.
- **Chemical doping:** Chemical doping in the separation membranes of LSBs plays a crucial role in altering the electronic structure of the host material to optimize the binding ability of LiPSs. Non-metal doping can enhance the chemical affinity to LiPSs through electron redistribution and the creation of active sites, while metal doping adjusts the Fermi level to influence surface charge distribution and improve LiPSs adsorption. Co-doping further synergizes these effects, enhancing the overall electrochemical stability and catalytic activity.
- **Strain engineering:** By applying tensile or compressive strain, the electronic structure of the material can be tuned, particularly by upshifting the d-band center in late transition metals, which strengthens the interaction with LiPSs. In late transition metals, tensile strain causes band narrowing and an upshift of the d-band center, while early transition metals show the opposite effect.
- **Crystal phase regulation:** Crystal phase regulation can enhance surface reactivity by modulating electronic properties and surface structures, which increases the density of dangling bond states and surface energy. This improvement leads to more active sites for LiPSs binding and conversion, ultimately boosting catalytic activity and facilitating faster $Li^+$ diffusion and SRR.
- **Disordered system:** Disordered systems, such as amorphous materials, can increase surface defects, dangling bonds, and broaden d-orbital electron distributions to improve the chemical affinity for LiPSs and catalyze conversion. However, while these systems are inherently unstable and require stabilization through surface passivation or heterostructure design to maintain long-term performance in the separation membranes of LSBs.
- **Hybrid materials:** Hybrid structures can effectively combine materials with complementary properties, such as high polarity and electrical conductivity materials. The incorporation of polymers or inorganic materials with high elastic modulus increases mechanical strength and thermal stability, preventing dendrite growth. Furthermore, introducing interfacial states and built-in electric fields enhances electron transfer and accelerates $Li^+$ transport, resulting in improved electrochemical kinetics.

## 5.2 Prospective

In designing separator membranes for LSBs, optimizing surface energy, active site configurations, and structural stability is crucial for enhancing LiPSs trapping and conversion efficiency. Effective modification strategies must balance catalytic activity with stability, focusing on selecting optimal active site configurations, preventing excessive LiPSs adsorption, and avoiding catalyst poisoning due to over-binding.

Separator membrane catalysts should focus on optimizing surface energy, dangling bond states, and the DOS near the Fermi level to enhance interaction with LiPSs. Dangling bonds at unsaturated coordination sites provide active sites for LiPSs adsorption and conversion, with the d/p-band centers modulating this interaction.

Several challenges persist, particularly the instability of high-energy states and excessive adsorption. High surface energy enhances the binding ability of LiPSs. However, it compromises stability, leading to capacity fade and impairing cycling performance due to structural damage or blocked Li$^+$ transport. While strategies like surface passivation and hybrid materials improve stability for materials, further advanced approaches are needed to balance binding affinity with stability, preventing catalyst poisoning and ensuring efficient LiPSs conversion during cycling.

Advanced characterization techniques and computational methods, such as DFT calculations, play a pivotal role in understanding and optimizing the performance of separation membranes for LSBs. These tools allow for detailed analysis of material properties, surface energy interactions, and the prediction of binding affinities at the atomic scale. Furthermore, in-situ techniques can provide insights into the dynamic behavior of LiPSs during cycling, enabling researchers to fine-tune the design and functionality of separator membranes in real-time. By guiding the design and fabrication of separator membranes, these discussions aim to accelerate the transformation of LSBs technology from basic research to practical applications.

*Acknowledgements*

This work is supported by the National Natural Science Foundation of China (52172228), and the Natural Science Foundation of Fujian Province (2024J01475 and 2023J05127). This report was proposed and convinced by Liangxu Lin. All other co-authors provided assistance and suggestions on the writing and presentation.

# TOC

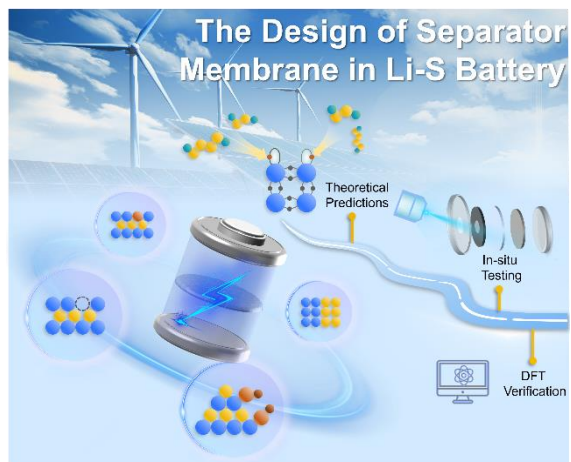

This report examines recent advancements in the rational design of separator membranes for lithium-sulfur batteries (LSBs), focusing on critical challenges related to surface binding and energy interactions in materials science. We aim to enhance the performance and stability of LSBs by optimizing separator membrane functionalization, focusing on the interplay between surface energy, binding affinities, and structural stability to inform future catalyst design for improved efficiency and durability.

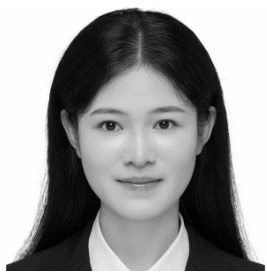

**Shuyu Cheng** is an MSc student at Fujian Normal University. She received her BSc degree from the Sichuan Normal University in 2021. Her research focuses on the energy storage technique for separator membranes modification of lithium-sulfur batteries.

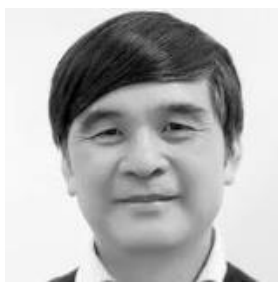

**Shaowei Zhang** is currently a Full Professor and Royal Society Industry Fellow at University of Exeter. He is also Fellows of Institute of Materials, and Royal Society of Chemistry. He had worked for a long period at University of Sheffield, as EPSRC Advanced Fellow, Lecturer and Reader. His main research interests are in the processing, microstructures and properties of structural and functional materials. He has also an research interest on 2D mateirals and interfaces.

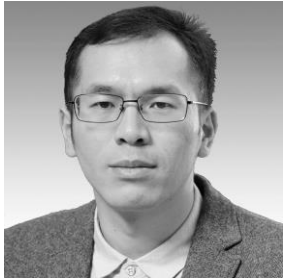

**Hongfang Du** is currently an Associate Professor at Fujian Normal University. He received his PhD degree in Clean Energy Science from Southwest University. He then worked as a research assistant at Nanyang Technological University and moved to Northwestern Polytechnical University as an assistant professor. His research interests focus on flexible electronics and clean energy conversion systems, including electrochemical water splitting, nitrogen reduction, carbon dioxide reduction.

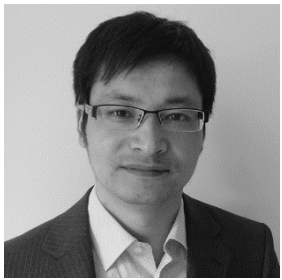

**Liangxu Lin** is currently a Full Professor at Fujian Normal University. He completed his PhD in Engineering Materials at The University of Sheffield at the end of 2013. After two postdoctoral periods at the University of Exeter, he joined Wuhan University of Science and Technology as a Professor, and the University of Wollongong as a Vice-Chancellor Fellow. His research focuses on materials interfaces and two-dimensional nanomaterials for electrochemical energy storage/conversion and catalysis.